\documentclass{article}
\usepackage{spconf,amsmath,graphicx,hyperref,booktabs,makecell,caption,xcolor}
\title{LTA-L2S: Lexical Tone-Aware Lip-to-Speech Synthesis for Mandarin with Cross-Lingual Transfer Learning}
\name{Kang Yang$^{1,2}$ \qquad Yifan Liang$^{3}$ \qquad Fangkun Liu$^{3}$ \qquad Zhenping Xie$^{{1,2,\dagger}}$\thanks{$^{\dagger}$Zhenping Xie is the corresponding author} \qquad Chengshi Zheng$^{3}$}
\address{$^{1}$ Jiangsu Key University Laboratory of Software and Media Technology under Human-Computer \\
Cooperation, Jiangnan University, Wuxi, China \\
$^{2}$ School of Artificial Intelligence and Computer Science, Jiangnan University, Wuxi, China \\
$^{3}$ Institute of Acoustics Chinese Academy of Science, Beijing, China }

\begin{document}
\ninept
\maketitle
\begin{abstract}
Lip-to-speech (L2S) synthesis for Mandarin is a significant challenge, hindered by complex viseme-to-phoneme mappings and the critical role of lexical tones in intelligibility. To address this issue, we propose Lexical Tone-Aware Lip-to-Speech (LTA-L2S). To tackle viseme-to-phoneme complexity, our model adapts an English pre-trained audio-visual self-supervised learning (SSL) model via a cross-lingual transfer learning strategy. This strategy not only transfers universal knowledge learned from extensive English data to the Mandarin domain but also circumvents the prohibitive cost of training such a model from scratch. To specifically model lexical tones and enhance intelligibility, we further employ a flow-matching model to generate the F0 contour. This generation process is guided by ASR-fine-tuned SSL speech units, which contain crucial suprasegmental information. The overall speech quality is then elevated through a two-stage training paradigm, where a flow-matching postnet refines the coarse spectrogram from the first stage. Extensive experiments demonstrate that LTA-L2S significantly outperforms existing methods in both speech intelligibility and tonal accuracy.
\end{abstract}


\begin{figure*}[htp]
  \centering
  \includegraphics[width=0.90\textwidth]{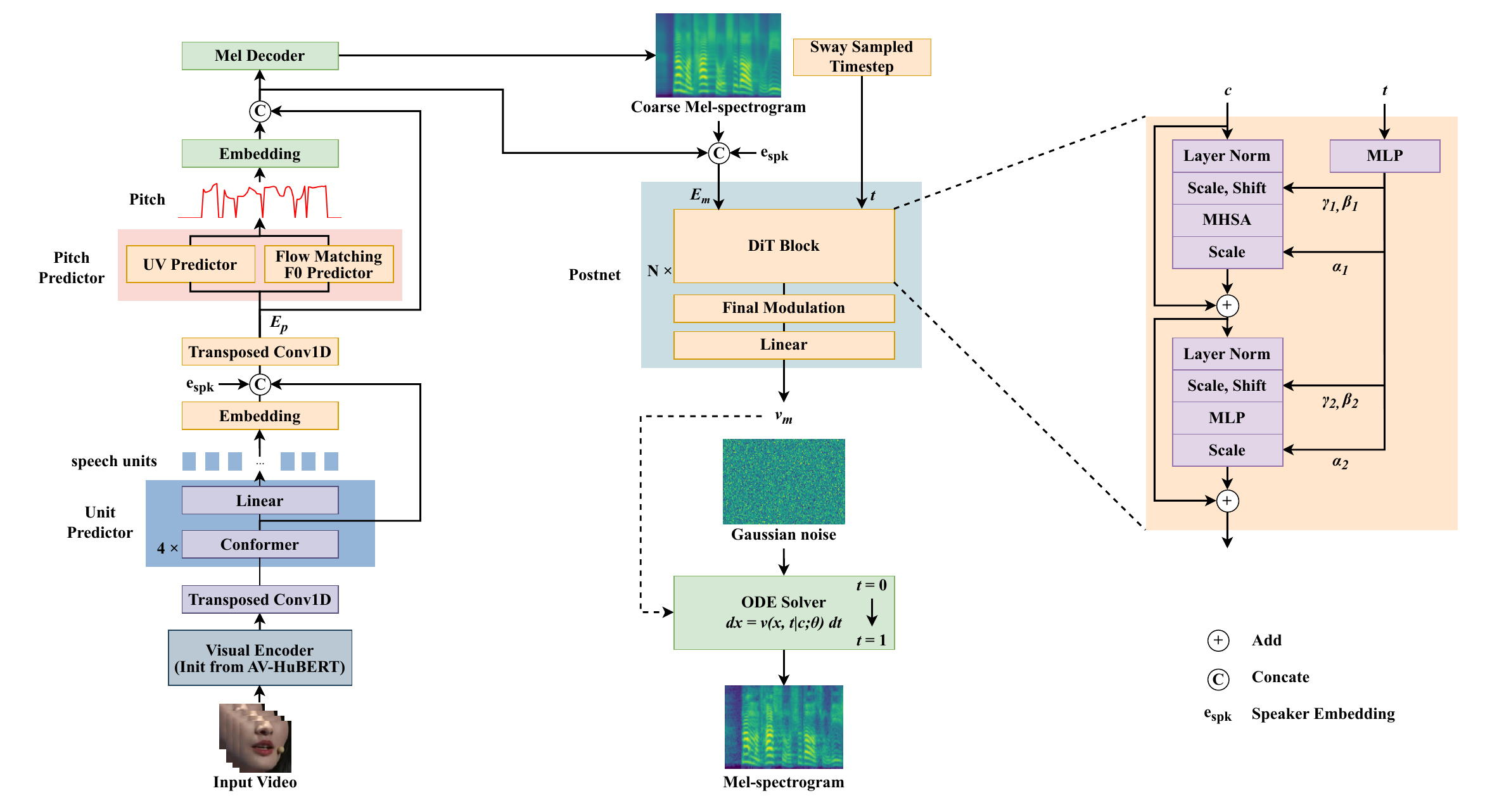}
  \captionsetup{justification=raggedright} 
  \caption{
     The overall architecture of our proposed LTA-L2S model. The left panel illustrates the main synthesis network, which processes visual features to generate a coarse mel-spectrogram. This coarse spectrogram is subsequently refined by the flow-matching postnet shown in the middle panel. The internal structure of the postnet's DiT block is detailed on the right.
      }
  \vspace{-0.5cm}
  \label{fig:main}
\end{figure*}
\section{Introduction}
\label{sec:intro}
{Human speech perception is an inherently multi-modal process, integrating both auditory and visual cues for robust comprehension. The influence of visual information, particularly lip movements, becomes critical in acoustically challenging environments, motivating the development of Lip-to-Speech (L2S) synthesis. L2S is the task of reconstructing intelligible speech from only lip movements, with potential applications ranging from restoring speech for individuals with vocal impairments \cite{hegde2023towards} to enhancing communication in noisy settings and dubbing silent films. Propelled by advances in deep learning, L2S systems have made substantial progress, moving closer to practical deployment.\par

The evolution of L2S ``in-the-wild'', driven by large-scale datasets \cite{afouras2018lrs3}, has seen significant progress. Foundational issues of poor intelligibility and unnaturalness have been largely overcome in English L2S through the integration of powerful pre-trained visual front-ends \cite{hegde2023towards, yemini2023lipvoicer, choi2025v2sflow, liang2025naturall2s, kim2025faces}, leveraging semantic supervision \cite{kim2023lip, choi2023intelligible, hsu2023revise, kim2024let}, and employing advanced generative frameworks \cite{yemini2023lipvoicer, choi2025v2sflow, liang2025naturall2s, kim2025faces, kim2024let, liang2025lightl2s}. However, developing an effective L2S system for Mandarin still remains a nascent field with substantial room for further improvement.\par
The task of L2S synthesis for Mandarin is impeded by two primary factors. The first one stems from Mandarin's large phoneme inventory, resulting in a more complex and ambiguous viseme-to-phoneme mapping. This ambiguity is particularly pronounced at the segmental level. For instance, due to Mandarin's initial-final syllable structure, many distinct initials combined with a common final (e.g., /zhi/, /chi/, /shi/, /ri/) produce visually similar or identical lip movements. Furthermore, visual cues alone are often insufficient to differentiate between crucial consonant classes like alveolar and retroflex. These segmental ambiguities present a formidable obstacle. The other factor is at the suprasegmental level, Mandarin's lexical tones, which can be influenced by F0 pattern, vowel duration, and amplitude \cite{tseng1986lexical}, are essential for distinguishing word meaning. However, these tonal variations are difficult to recognize from visual information alone, creating a significant barrier to generating intelligible and natural-sounding Mandarin speech. }\par 

To address these above-mentioned specific challenges, we propose Lexical Tone-Aware L2S (LTA-L2S), a novel framework that makes two primary contributions. First, the framework tackles the complex viseme-to-phoneme mappings through a cross-lingual transfer learning strategy. This approach effectively leverages knowledge from English audio-visual data while reducing computational cost. Second, the synthesis core of the framework employs a specialized pitch predictor based on flow-matching generative model guided by ASR-fine-tuned SSL speech units that contain crucial suprasegmental information to precisely model lexical tones and improve intelligibility. We employ a flow-matching postnet to refine coarse mel-spectrogram, which ensures high naturalness and acoustic detail. Extensive evaluations on the CN-CVS dataset \cite{chen2023cn} demonstrate that our approach significantly outperforms existing methods in both speech intelligibility and tonal accuracy. Audio samples are available on our demo page\footnote{\url{https://yangkangjnu.github.io/LTA-L2S_demo/}}.

\section{Method}
\label{sec:method}
The overall framework of our proposed LTA-L2S model is illustrated in Fig. 1. The process begins with a visual encoder and a unit predictor that predict speech units from the silent lip video. Then, the pitch predictor, which consists of a unvoiced (UV) predictor for discrete UV label and a flow matching F0 predictor for generating the F0 contour, predicts the pitch mainly guided by the predicted speech units. Given the speech units and pitch information, the mel decoder then generates a coarse mel-spectrogram. Finally, the coarse mel-spectrogram is refined by a flow-matching postnet to enhance acoustic detail. The model is optimized using a two-stage training strategy. In the first stage, the entire network up to the mel decoder is trained to generate the coarse spectrogram. In the second stage, the parameters of the first-stage model are frozen, and only the postnet is trained to perform the refinement task. 

\subsection{Visual Encoder}
\label{ssec:visual_encoder} 
Our visual encoder employs a cross-lingual transfer learning strategy to leverage knowledge learned from large-scale English audio-visual datasets. Its architecture is based on the AV-HuBERT Base model \cite{shi2022learning}, comprising a 3D convolution layer, a ResNet-18 backbone, and a Transformer encoder. Inspired by previous work \cite{zhang2025target, zinonos2023learning} on Mandarin lip-reading, we initialize this encoder with weights pre-trained on 1759 hours of English data and subsequently fine-tune it on our Mandarin dataset during the first training stage. This approach leverages the high degree of viseme sharedness across languages, allowing us to transfer general knowledge to the Mandarin domain. Crucially, it circumvents the prohibitive computational cost of audio-visual self-supervised pre-training and enhances performance under limited data conditions. 

\subsection{Unit Predictor}
\label{ssec:unit_predictor}
The unit predictor is designed to generate a sequence of discrete speech units that serve as the primary content representation. 
Building on prior work \cite{de2024layer, shen2024encoding} showing that features from the ASR-fine-tuned SSL speech model yield superior tone classification performance compared to the non-fine-tuned model, we extract the target units by applying k-means clustering to the 9th-layer features of an ASR-fine-tuned wav2vec 2.0 model (w2v2-ft) \cite{baevski2020wav2vec, lu2023context}. This layer was chosen because its features achieve the highest performance in tone classification, suggesting they contain the most abundant suprasegmental information. To predict these units, the predictor adopts the Conformer architecture \cite{gulati2020conformer}, comprising four Conformer blocks, and a final linear layer. The output of visual encoder is temporally upsampled by a factor of two via a transposed 1D convolution to match the speech units frame rate, and then passed to the unit predictor. The model is trained by minimizing a cross-entropy (CE) loss ($\mathcal{L}_{unit}$) with a label smoothing factor of 0.1 between its predictions and the target units. This loss is defined as:
\begin{equation}
  \mathcal{L}_{unit} = - \frac{1}{T_{u}} \sum_{i=1}^{T_{u}} \sum_{j=1}^{C} q_{i,j} \log \left( \frac{\exp(U_{i,j})}{\sum_{k=1}^{C} \exp(U_{i,k})} \right), \label{eq_unitloss}
\end{equation}
where $T_{u}$ is the sequence length of the speech units, $C$ is the total number of clusters, $U_{i,j}$ denotes the predicted logits for the j-th unit of the i-th frame, and $q_{i,j}$ represents the label-smoothed target distribution.

\subsection{Pitch Predictor}
\label{ssec:pitch_predictor}
The pitch predictor comprises an F0 predictor for generating the continuous F0 contour and a UV predictor for determining the discrete UV labels. 
We employ the \texttt{YAAPT} algorithm \cite{kasi2002yet} to extract the pitch sequence from raw audio, which is then decomposed into a continuous F0 contour and discrete UV labels. 
To accurately model the complex variations in F0 contours, we adopt a flow-matching approach, inspired by its success in the demanding domain of singing voice synthesis \cite{guo2025techsinger}. 
We define a probability path by linearly interpolating between the target F0 contour $x_1$ and Gaussian noise $x_0$. 
An intermediate sample on this path is given by $x_t=(1-t)x_0+tx_1$ for $t\in[0,1]$. 
Our goal is to train a vector field estimator $\theta$ to predict the vector field $u=x_1-x_0$ with the corresponding condition $c$, which is the combination features $E_p$, generated by upsampling the combination of hidden visual features, speech unit embeddings, and speaker embeddings through a transposed 1D convolution. 
Consistent with prior works \cite{choi2023intelligible, mira2022svts}, 
the speaker embeddings are extracted from a pre-trained speaker verification model\footnote{\url{https://github.com/CorentinJ/Real-Time-Voice-Cloning}}.
We employ a Diffusion Transformer (DiT) \cite{peebles2023scalable} as the estimator's backbone and optimize it using the Rectified Flow Matching (RFM) \cite{liu2022flow} loss: 
\begin{equation}
    \mathcal{L}_{f0} = ||v(x_t,t|c; \theta) - (x_1 - x_0)||^2, \label{eq_f0loss}
\end{equation}
where $v(x_t,t|c; \theta)$ is the estimated vector field. 
During inference, the final F0 counter is generated by sampling an initial point $x_0 \sim \mathcal{N}(0, I)$ and solving the corresponding ordinary differential equation (ODE) from $t=0$ to $t=1$ via the Euler method: 
\begin{equation}
    x_{t + \frac{1}{N}} = x_t + \frac{1}{N} \hat{v}_{\theta}(x_t, c, t),\label{eq_ode}
\end{equation}
where \textit{N} is the number of function evaluations (NFE).
The UV predictor consists of a stack of five 1D convolution layers followed by a linear layer.  
It is trained to predict the binary UV labels by minimizing a binary cross-entropy (BCE) loss, denoted as $\mathcal{L}_{uv}$. The final pitch is obtained by applying the UV mask to the continuous F0 contour, setting the unvoiced frames to zero and retaining the voiced values. 

\subsection{Mel Decoder}
\label{ssec:mel_decoder}
The final component of the main network in the first training stage is the mel decoder, which is based on the FastSpeech2 \cite{ren2020fastspeech} decoder architecture, comprising a stack of six FFT (Feed Forward Transformer) blocks and a final linear layer. 
It is trained to reconstruct the mel-spectrogram from a combination of hidden visual features, speech unit embeddings, speaker embeddings, and pitch embeddings.
This reconstruction process is optimized by minimizing an L1 loss, 
which we denote as $\mathcal{L}_{mel}$:
\begin{equation}
  \mathcal{L}_{mel} = \frac{1}{T_{m}} \sum_{i=1}^{T_{m}} \left\| M_{i} - \hat{M_{i}} \right\|_1, \label{eq_melloss}
\end{equation}  
where $T_{m}$ is the total number of frames in the mel-spectrogram, 
and $M_{i}$ and $\hat{M_{i}}$ are the ground-truth and predicted mel-spectrogram for the i-th frame, 
respectively.

The overall objective for the fisrt training stage $\mathcal{L}_{stage1}$ is a weighted sum of the four above-defined component losses: 
\begin{equation}
    \mathcal{L}_{stage1}=\lambda_{unit}\mathcal{L}_{unit} + \lambda_{uv}\mathcal{L}_{uv} + \lambda_{f0}\mathcal{L}_{f0} + \lambda_{mel}\mathcal{L}_{mel} \label{eq_stage1}
\end{equation}  
where $\lambda_{unit}$, $\lambda_{uv}$, $\lambda_{f_0}$ and $\lambda_{mel}$ are the respective loss weights. 
Based on empirical evaluation, we set these weights to $\lambda_{unit}=0.1$, and $\lambda_{uv}=\lambda_{f_0}=\lambda_{mel}=1.0$.

\subsection{Postnet}
\label{ssec:postnet}
To overcome the limitations of the L1 loss, which often yields overly smoothed spectrograms, we introduce a second training stage where a flow-matching postnet refines the coarse output. This postnet employs a Diffusion Transformer (DiT) backbone and is optimized with a RFM loss, $\mathcal{L}_{postnet}$, analogous to that of the F0 predictor (Eq. \ref{eq_f0loss}). During this stage,the main network trained in the first stage is frozen, and the postnet is trained to perform this refinement conditioned on the combined representation of the coarse spectrogram and the mel decoder's input ($E_m$). The final, fine-grained mel-spectrogram is then synthesized during inference by applying this trained postnet to the coarse output via an ODE solver.

\section{Experimental Settings}
\label{sec:experimental_settings}
\subsection{Datasets}
\label{ssec:datasets}
We evaluated LTA-L2S on the speech subset of the CN-CVS dataset \cite{chen2023cn}, which is the first publicly available large-scale Mandarin audio-visual dataset widely used for lip-reading task. This speech subset comprises 273.4 hours of audio-visual data from 2529 speakers. We followed the original split proposed in CN-CVS for unseen speaker scenarios and excluded non-Mandarin sentences. Furthermore, we re-transcribed the text labels for the test set using FunASR model\footnote{\url{https://github.com/modelscope/FunASR}}.

\subsection{Implementation Details}
\label{ssec:implementation_details}
For data preprocessing, we first extract 68 facial landmarks using dlib \cite{king2009dlib} to align each frame. The aligned frames are then cropped to a $96 \times 96$ patch centered on the lips and converted to grayscale. 
The corresponding audio is sampled to 16 kHz, 
and mel spectrogram is extracted using 80 mel filter banks with 40ms window and 10ms hop size.
For data augmentation, we follow the methods proposed in previous work\cite{mira2022svts}.

The model's architectural and training parameters are configured as follows. The discrete speech units are derived by using k-means algorithm with 2000 clusters. The DiT backbones for the F0 predictor and post-net share a configuration of 8 attention heads, a 512-dimensional embedding, and a 1024-dimensional feed-forward network, but differ in depth with 8 and 12 layers, respectively. We train the model for 100 epochs per stage on two NVIDIA 4090 GPUs with a batch size of 16 per GPU, employing a teacher-forcing strategy throughout. The model is optimized using AdamW \cite{loshchilov2017decoupled} with a learning rate that peaks at 1e-4, governed by a 10-epoch linear warm-up and a subsequent cosine decay schedule. For inference, both the F0 predictor and the postnet adopt the the sway sampling startegy \cite{chen2024f5} with NFE of 24. To convert the mel-spectrograms to waveforms, we employ a HiFi-GAN vocoder \cite{kong2020hifi} pre-trained on the CN-CVS speech dataset.

\subsection{Baseline Systems}
\label{ssec:baseline_systems}
Our method is compared with several state-of-the-art methods: VCA-GAN \cite{kim2021lip}, SVTS \cite{mira2022svts}, Multi-task \cite{kim2023lip}, Intelligible \cite{choi2023intelligible} and LTBS \cite{kim2024let}. 
In terms of implementation, we follow the official code of VCA-GAN and Intelligible, 
and for SVTS and LTBS, we reproduce them according to their original papers. 
\subsection{Evaluation Metrics}
\label{ssec:evaluation_metrics}
Our model's performance is evaluated using both objective and subjective metrics. For objective analysis, we measure speech quality with DNSMOS \cite{reddy2021dnsmos} and STOI-Net \cite{zezario2020stoi}, and speaker similarity via Speaker Embedding Cosine Similarity (SECS), using embeddings from Resemblyzer \cite{wan2018generalized}. Intelligibility and tonal accuracy are assessed using Character Error Rate (CER) and Tone Error Rate (TER), respectively. Both are computed from FunASR transcriptions post-processed by the g2pM library\footnote{\url{https://github.com/kakaobrain/g2pm}}. For subjective evaluation, we conducted a Mean Opinion Score (MOS) listening test where 15 native Mandarin speakers rated 30 samples per model on a 5-point scale for naturalness, intelligibility, and speaker similarity.

\begin{table}[t]
  \renewcommand{\arraystretch}{1.2}
  \renewcommand{\tabcolsep}{1.0mm}
  \caption{
      Objective evaluation results on the CN-CVS dataset.
  }
  \vspace{-0.2cm}
  \centering
  \begin{tabular}{l ccccc}
      \Xhline{3\arrayrulewidth}
      Method & STOI-Net$\uparrow$ & DNSMOS$\uparrow$ & CER$\downarrow$ & TER$\downarrow$ & SECS$\uparrow$ \\
      \cmidrule(l{2pt}r{2pt}){1-1} \cmidrule(l{2pt}r{2pt}){2-6}
      VCA-GAN \cite{kim2021lip} & 0.334 & 2.206 & 96.5 & 78.1 & 0.523 \\
      SVTS \cite{mira2022svts} & 0.651 & 2.292 & 87.4 & 59.2 & 0.587 \\
      Multi-task \cite{kim2023lip} & 0.443 & 2.240 & 84.3 & 71.0 & 0.547 \\
      Intelligible \cite{choi2023intelligible} & 0.766 & 2.432 & 74.9 & 50.3 & 0.667 \\
      LTBS \cite{kim2024let} & \textbf{0.909} & \textbf{3.291} & 87.3 & 56.4 & 0.782 \\
      \textbf{LTA-L2S} & 0.907 & 3.200 & \textbf{61.2} & \textbf{40.9} & \textbf{0.828} \\
      Ground Truth & 0.908 & 3.302 & -- & -- & -- \\
      \Xhline{3\arrayrulewidth}
  \end{tabular}
  \vspace{-0.2cm}
  \label{tabel:1}
\end{table}

\begin{table}[t]
  \renewcommand{\arraystretch}{1.2}
  \renewcommand{\tabcolsep}{1.0mm}
  \caption{
      Subjective evaluation results on the CN-CVS dataset.
  }
  \vspace{-0.2cm}
  \centering
  \begin{tabular}{l ccc}
    \Xhline{3\arrayrulewidth}
      Method & Naturalness$\uparrow$ & Intelligibility$\uparrow$  & Similarity$\uparrow$ \\
      \cmidrule(l{2pt}r{2pt}){1-1} \cmidrule(l{2pt}r{2pt}){2-4}
      VCA-GAN & 1.60$\pm$0.44 & 1.82$\pm$0.60 & 1.71$\pm$0.53  \\
      SVTS & 1.70$\pm$0.26 & 2.04$\pm$0.50 & 1.90$\pm$0.43 \\
      Multi-task & 1.61$\pm$0.38 & 2.00$\pm$0.62 & 1.82$\pm$0.51 \\
      Intelligible & 2.36$\pm$0.46 & 2.49$\pm$0.51 & 2.39$\pm$0.58 \\
      LTBS & 2.89$\pm$0.81 & 2.17$\pm$0.53 & 2.51$\pm$0.70 \\
      \textbf{LTA-L2S} & \textbf{3.95$\pm$0.34} & \textbf{3.88$\pm$0.30} & \textbf{3.89$\pm$0.42} \\
      Ground Truth & 4.48$\pm$0.16 & 4.54$\pm$0.15 & --\\
      \Xhline{3\arrayrulewidth}
  \end{tabular}
  \label{tabel:2}
  \vspace{-0.2cm}
\end{table}

\section{Experimental Results}
\label{sec:experimental_results}
\subsection{Objective Evaluation}
\label{ssec:objective_evaluation}
As reported in Table 1,our proposed LTA-L2S model demonstrates comprehensive superiority over baseline methods on the CN-CVS dataset. This advantage is most pronounced in intelligibility, where our model achieves state-of-the-art results with the lowest CER (61.2\%) and TER(40.9\%), underscoring its advanced capability in capturing both semantic content and crucial lexical tones. For speech quality, our method obtains competitive STOI-Net (0.907) and DNSMOS (3.200) scores, ranking second only to LTBS but substantially ahead of other approaches and only marginally below the ground truth. Meanwhile, the proposed LTA-L2S also achieves the highest speaker similarity with a top-ranking SECS score of 0.828, confirming its excellent capability in preserving speaker identity.

\subsection{Subjective Evaluation}
\label{ssec:subjective_evaluation}
The results of subjective evaluation, presented in Table 2, further reinforce our model's superiority. LTA-L2S was consistently preferred over all baselines across all metrics: naturalness, intelligibility, and speaker similarity. A key finding emerges when comparing objective and subjective results for the LTBS model. Notably, while LTBS achieved high scores in objective quality metrics, its poor intelligibility appears to have negatively impacted its perceived naturalness in subjective tests, a phenomenon that highlights the importance of intelligibility in holistic speech perception. In contrast, the superior MOS scores of our model in all categories underscore its ability to strike an effective balance, accurately reconstructing speech content from Mandarin lip videos while faithfully preserving the speaker’s vocal identity and achieving high naturalness.

\subsection{Ablation Study}
\label{ssec:ablation_study}
To isolate the contribution of each proposed component, we conducted a comprehensive ablation study, with the results summarized in Table 3. The analysis yields several key insights. First, removing the postnet caused a significant drop in DNSMOS, confirming its crucial role in enhancing speech naturalness. This result highlights its ability to refine fine-grained acoustic details and counteract the over-smoothing inherent in simple L1 loss optimization. Second, generating F0 under the optimization of a simple L1 loss instead of our proposed method resulted in higher CER and TER, affirming the effectiveness of our flow-matching approach. Furthermore, removing the entire pitch predictor led to a marked decline in both STOI-Net and DNSMOS scores, underscoring the foundational role of pitch in improving speech quality. Finally, ablating the supervision from speech units caused substantial performance degradation, confirming their importance for the Mandarin lip-to-speech system.

Finally, we conducted ablation studies to isolate the contributions of the cross-lingual transfer learning strategy and the discrete speech units from w2v2-ft model, and the results are presented in Table 4. First, the absence of initialization with AV-HuBERT Base weitghts for the visual encoder led to a substantial increase in both CER and PER. This result validates the effectiveness of our cross-lingual transfer learning strategy. To assess the efficacy of discrete speech units derived from w2v2-ft model, we replaced them with discrete units derived from K-means clustering (k=2000) applied to the features from the 9th layer of a pre-trained HuBERT model, which is frequently used in English L2S tasks. This modification resulted in a notable increase in both CER and TER, confirming that the speech units from the w2v2-ft model can effectively provide both semantic supervision and supplementary lexical tone information for the Mandarin L2S task.

\begin{table}[t]
  \renewcommand{\arraystretch}{1.2}
  \renewcommand{\tabcolsep}{1.0mm}
  \caption{
      Ablation study of model components on the CN-CVS dataset.
  }
  \vspace{-0.2cm}
  \centering
  \begin{tabular}{l ccccc}
    \Xhline{3\arrayrulewidth}
      Method & STOI-Net$\uparrow$ & DNSMOS$\uparrow$ & CER$\downarrow$ & TER$\downarrow$ & SECS$\uparrow$  \\ 
      \cmidrule(l{2pt}r{2pt}){1-1} \cmidrule(l{2pt}r{2pt}){2-6}
      LTA-L2S & \textbf{0.907} & \textbf{3.200} & 61.2 & 40.9 & \textbf{0.828} \\
      \cmidrule(l{2pt}r{2pt}){1-1} \cmidrule(l{2pt}r{2pt}){2-6}
      \textit{w/o} postnet & 0.902 & 2.942 & \textbf{61.0} & \textbf{40.7} & 0.821 \\
      ~~\textit{w/o} f0 cfm & 0.906 & 2.972 & 63.2 & 42.0 & 0.817 \\
      ~~~~\textit{w/o} pitch & 0.877 & 2.731 & 61.5 & 40.9 & 0.811 \\
      ~~~~~~\textit{w/o} unit & 0.771 & 2.395 & 69.2 & 46.2 & 0.660 \\
      \Xhline{3\arrayrulewidth}
  \end{tabular}
  \vspace{-0.2cm}
  \label{tabel:3}
\end{table}

\begin{table}[t]
  \renewcommand{\arraystretch}{1.2}
  \renewcommand{\tabcolsep}{1.0mm}
  \caption{
      Ablation study of cross-lingual transfer learning and discrete speech units on the CN-CVS dataset
  }
  \vspace{-0.2cm}
  \centering
  \begin{tabular}{l ccccc}
    \Xhline{3\arrayrulewidth}
      Method & STOI-Net$\uparrow$ & DNSMOS$\uparrow$ & CER$\downarrow$ & TER$\downarrow$ & SECS$\uparrow$  \\ 
      \cmidrule(l{2pt}r{2pt}){1-1} \cmidrule(l{2pt}r{2pt}){2-6}
      LTA-L2S & \textbf{0.907} & 3.200 & \textbf{61.2} & \textbf{40.9} & 0.828 \\
      \cmidrule(l{2pt}r{2pt}){1-1} \cmidrule(l{2pt}r{2pt}){2-6}
      Random Init. & 0.903 & 3.174 & 74.6 & 49.1 & 0.822 \\
      HuBERT Unit & 0.905 & \textbf{3.206} & 64.4 & 42.8 & \textbf{0.831} \\
      \Xhline{3\arrayrulewidth}
  \end{tabular}
  \label{tabel:4}
  \vspace{-0.2cm}
\end{table}

\section{Conclusion}
\label{sec:conclusion}
In this paper, we proposed LTA-L2S, a novel framework for Mandarin Lip-to-Speech synthesis. The framework enhances semantic accuracy via cross-lingual transfer learning, addresses tonal intelligibility by guiding a flow-matching F0 predictor with suprasegmental speech units, and elevates overall naturalness with a dedicated flow-matching postnet for mel-spectrogram refinement. Our experimental results confirm the superiority of LTA-L2S over previous methods in both speech quality and intelligibility. For future work, we aim to extend this framework to other Chinese dialects and accents. 

\vfill\pagebreak
{

\bibliographystyle{IEEEbib}
\bibliography{references}

\begin{thebibliography}{10}

\bibitem{hegde2023towards}
S.~Hegde, R.~Mukhopadhyay, C.~V. Jawahar, and V.~Namboodiri,
\newblock ``{Towards Accurate Lip-to-Speech Synthesis in-the-Wild},''
\newblock in {\em ACM MM}, 2023, pp. 5523--5531.

\bibitem{afouras2018lrs3}
T.~Afouras, J.~S. Chung, and A.~Zisserman,
\newblock ``{LRS3-TED: A Large-Scale Dataset for Visual Speech Recognition},''
\newblock {\em arXiv preprint arXiv:1809.00496}, 2018.

\bibitem{yemini2023lipvoicer}
Y.~Yemini, A.~Shamsian, L.~Bracha, S.~Gannot, and E.~Fetaya,
\newblock ``{LipVoicer: Generating Speech from Silent Videos Guided by Lip Reading},''
\newblock {\em arXiv preprint arXiv:2306.03258}, 2023.

\bibitem{choi2025v2sflow}
J.~Choi, J.-H. Kim, J.~Li, J.~S. Chung, and S.~Liu,
\newblock ``{V2SFlow: Video-to-Speech Generation with Speech Decomposition and Rectified Flow},''
\newblock in {\em ICASSP}, 2025, pp. 1--5.

\bibitem{liang2025naturall2s}
Y.~Liang, F.~Liu, A.~Li, X.~Li, and C.~Zheng,
\newblock ``{NaturalL2S: End-to-End High-Quality Multispeaker Lip-to-Speech Synthesis with Differential Digital Signal Processing},''
\newblock {\em arXiv preprint arXiv:2502.12002}, 2025.

\bibitem{kim2025faces}
J.-H. Kim, J.~Choi, J.~Kim, C.~Jung, and J.~S. Chung,
\newblock ``{From Faces to Voices: Learning Hierarchical Representations for High-Quality Video-to-Speech},''
\newblock in {\em CVPR}, 2025, pp. 15874--15884.

\bibitem{kim2023lip}
M.~Kim, J.~Hong, and Y.~M. Ro,
\newblock ``{Lip-to-Speech Synthesis in the Wild with Multi-Task Learning},''
\newblock in {\em ICASSP}, 2023, pp. 1--5.

\bibitem{choi2023intelligible}
J.~Choi, M.~Kim, and Y.~M. Ro,
\newblock ``{Intelligible Lip-to-Speech Synthesis with Speech Units},''
\newblock {\em arXiv preprint arXiv:2305.19603}, 2023.

\bibitem{hsu2023revise}
W.-N. Hsu, T.~Remez, B.~Shi, J.~Donley, and Y.~Adi,
\newblock ``{Revise: Self-Supervised Speech Resynthesis with Visual Input for Universal and Generalized Speech Regeneration},''
\newblock in {\em CVPR}, 2023, pp. 18795--18805.

\bibitem{kim2024let}
J.~H. Kim, J.~Kim, and J.~S. Chung,
\newblock ``{Let There Be Sound: Reconstructing High Quality Speech from Silent Videos},''
\newblock in {\em AAAI}, 2024, vol.~38, pp. 2759--2767.

\bibitem{liang2025lightl2s}
Y.~Liang, K.~Yang, F.~Liu, A.~Li, X.~Li, and C.~Zheng,
\newblock ``{LightL2S: Ultra-Low Complexity Lip-to-Speech Synthesis for Multi-Speaker Scenarios},''
\newblock in {\em Interspeech}, 2025, pp. 3783--3787.

\bibitem{tseng1986lexical}
C.~Y. Tseng, D.~W. Massaro, and M.~M. Cohen,
\newblock ``{Lexical Tone Perception in Mandarin Chinese: Evaluation and Integration of Acoustic Features},''
\newblock {\em Linguistics, psychology, and the Chinese language}, pp. 91--104, 1986.

\bibitem{chen2023cn}
C.~Chen, D.~Wang, and T.~F. Zheng,
\newblock ``{CN-CVS: A Mandarin Audio-Visual Dataset for Large Vocabulary Continuous Visual to Speech Synthesis},''
\newblock in {\em ICASSP}, 2023, pp. 1--5.

\bibitem{shi2022learning}
B.~Shi, W.-N. Hsu, K.~Lakhotia, and A.~Mohamed,
\newblock ``{Learning Audio-Visual Speech Representation by Masked Multimodal Cluster Prediction},''
\newblock {\em arXiv preprint arXiv:2201.02184}, 2022.

\bibitem{zhang2025target}
J.-X. Zhang, T.~Mao, L.~Guo, J.~Li, and L.~Zhang,
\newblock ``{Target Speaker Lipreading by Audio–Visual Self-Distillation Pretraining and Speaker Adaptation},''
\newblock {\em Expert Syst. Appl.}, vol. 272, pp. 126741, 2025.

\bibitem{zinonos2023learning}
A.~Zinonos, A.~Haliassos, P.~Ma, S.~Petridis, and M.~Pantic,
\newblock ``{Learning Cross-Lingual Visual Speech Representations},''
\newblock in {\em ICASSP}, 2023, pp. 1--5.

\bibitem{de2024layer}
A.~de~la Fuente and D.~Jurafsky,
\newblock ``{A Layer-Wise Analysis of Mandarin and English Suprasegmentals in SSL Speech Models},''
\newblock {\em arXiv preprint arXiv:2408.13678}, 2024.

\bibitem{shen2024encoding}
G.~Shen, M.~Watkins, A.~Alishahi, and A.~Bisazza,
\newblock ``{Encoding of Lexical Tone in Self-Supervised Models of Spoken Language},''
\newblock {\em arXiv preprint arXiv:2403.16865}, 2024.

\bibitem{baevski2020wav2vec}
A.~Baevski, Y.~Zhou, A.~Mohamed, and M.~Auli,
\newblock ``{Wav2Vec 2.0: A Framework for Self-Supervised Learning of Speech Representations},''
\newblock {\em NeurIPS}, vol. 33, pp. 12449--12460, 2020.

\bibitem{lu2023context}
K.~H. Lu and K.~Y. Chen,
\newblock ``{A Context-Aware Knowledge Transferring Strategy for CTC-Based ASR},''
\newblock in {\em SLT}, 2023, pp. 60--67.

\bibitem{gulati2020conformer}
A.~Gulati, J.~Qin, C.-C. Chiu, N.~Parmar, Y.~Zhang, J.~Yu, W.~Han, S.~Wang, Z.~Zhang, and Y.~Wu,
\newblock ``{Conformer: Convolution-Augmented Transformer for Speech Recognition},''
\newblock {\em arXiv preprint arXiv:2005.08100}, 2020.

\bibitem{kasi2002yet}
K.~Kasi and S.~A. Zahorian,
\newblock ``{Yet Another Algorithm for Pitch Tracking},''
\newblock in {\em ICASSP}, 2002, vol.~1, pp. I--361.

\bibitem{guo2025techsinger}
W.~Guo, Y.~Zhang, C.~Pan, R.~Huang, L.~Tang, R.~Li, Z.~Hong, Y.~Wang, and Z.~Zhao,
\newblock ``{TechSinger: Technique Controllable Multilingual Singing Voice Synthesis via Flow Matching},''
\newblock in {\em AAAI}, 2025, vol.~39, pp. 23978--23986.

\bibitem{mira2022svts}
R.~Mira, A.~Haliassos, S.~Petridis, B.~W. Schuller, and M.~Pantic,
\newblock ``{SVTS: Scalable Video-to-Speech Synthesis},''
\newblock {\em arXiv preprint arXiv:2205.02058}, 2022.

\bibitem{peebles2023scalable}
W.~Peebles and S.~Xie,
\newblock ``{Scalable Diffusion Models with Transformers},''
\newblock in {\em ICCV}, 2023, pp. 4195--4205.

\bibitem{liu2022flow}
X.~Liu, C.~Gong, and Q.~Liu,
\newblock ``{Flow Straight and Fast: Learning to Generate and Transfer Data with Rectified Flow},''
\newblock {\em arXiv preprint arXiv:2209.03003}, 2022.

\bibitem{ren2020fastspeech}
Y.~Ren, C.~Hu, X.~Tan, T.~Qin, S.~Zhao, Z.~Zhao, and T.-Y. Liu,
\newblock ``{FastSpeech 2: Fast and High-Quality End-to-End Text to Speech},''
\newblock {\em arXiv preprint arXiv:2006.04558}, 2020.

\bibitem{king2009dlib}
D.~E. King,
\newblock ``{Dlib-ml: A Machine Learning Toolkit},''
\newblock {\em J. Mach. Learn. Res.}, vol. 10, pp. 1755--1758, 2009.

\bibitem{loshchilov2017decoupled}
I.~Loshchilov and F.~Hutter,
\newblock ``{Decoupled Weight Decay Regularization},''
\newblock {\em arXiv preprint arXiv:1711.05101}, 2017.

\bibitem{chen2024f5}
Y.~Chen, Z.~Niu, Z.~Ma, K.~Deng, C.~Wang, J.~Zhao, K.~Yu, and X.~Chen,
\newblock ``{F5-TTS: A Fairytaler That Fakes Fluent and Faithful Speech with Flow Matching},''
\newblock {\em arXiv preprint arXiv:2410.06885}, 2024.

\bibitem{kong2020hifi}
J.~Kong, J.~Kim, and J.~Bae,
\newblock ``{HiFi-GAN: Generative Adversarial Networks for Efficient and High Fidelity Speech Synthesis},''
\newblock {\em NeurIPS}, vol. 33, pp. 17022--17033, 2020.

\bibitem{kim2021lip}
M.~Kim, J.~Hong, and Y.~M. Ro,
\newblock ``{Lip to Speech Synthesis with Visual Context Attentional GAN},''
\newblock {\em NeurIPS}, vol. 34, pp. 2758--2770, 2021.

\bibitem{reddy2021dnsmos}
C.~K.~A. Reddy, V.~Gopal, and R.~Cutler,
\newblock ``{DNSMOS: A Non-Intrusive Perceptual Objective Speech Quality Metric to Evaluate Noise Suppressors},''
\newblock in {\em ICASSP}, 2021, pp. 6493--6497.

\bibitem{zezario2020stoi}
R.~E. Zezario, S.-W. Fu, C.-S. Fuh, Y.~Tsao, and H.-M. Wang,
\newblock ``{STOI-Net: A Deep Learning Based Non-Intrusive Speech Intelligibility Assessment Model},''
\newblock in {\em APSIPA ASC}, 2020, pp. 482--486.

\bibitem{wan2018generalized}
L.~Wan, Q.~Wang, A.~Papir, and I.~L. Moreno,
\newblock ``{Generalized End-to-End Loss for Speaker Verification},''
\newblock in {\em ICASSP}, 2018, pp. 4879--4883.

\end{thebibliography}
}

\end{document}